\documentstyle[12pt,aasms4]{article}

\def\rerg{\rm erg}
\def\rs{\rm s}
\def\rs1{\rm s^{-1}}

\def\cmm2{\rm cm^{-2}}
\def\deg{\rm ^{\circ}}
\def\flux{\rerg\ \rcm2\ \rs1}

\def\zeta{A_{Fe}}

\def\rerg{\rm erg}
\def\rs{\rm s}
\def\rcm{\rm cm}

\def\flux{\rm erg\ cm^{-2}\ s^{-1}}
\def\fluence{\rm erg\ cm^{-2}}

\def\csq{\chi^2}

\def\ltsim{\lower 0.5ex\hbox{${}\buildrel<\over\sim{}$}}
\def\gtrsim{\lower 0.5ex\hbox{${}\buildrel>\over\sim{}$}}

\def\etal{{\it et al. }}
\def\apj{ApJ}
\def\apjl{ApJ}

\def\nature{Nature }
\def\nat{ Nature }
\def\aa{A\&A}

\def\aas{ A\&AS}
\def\mnras{MNRAS}

\begin{document}

\title{
The X-ray afterglow of GRB000926 observed by BeppoSAX and Chandra:
a mildly collimated fireball in a dense medium ?}

\slugcomment{Accepted for publication in ApJ}

\author{L. Piro\altaffilmark{1},
G. Garmire\altaffilmark{2}, M. R. Garcia\altaffilmark{3}, L. A.
Antonelli\altaffilmark{4} E. Costa\altaffilmark{1}, M.
Feroci\altaffilmark{1}, D. A. Frail\altaffilmark{5},
F. Harrison\altaffilmark{6},    K. Hurley\altaffilmark{7},
P. M\'esz\'aros\altaffilmark{2}, E.
Waxman\altaffilmark{8} }

\altaffiltext{1}{Istituto Astrofisica Spaziale, C.N.R., Via Fosso
del Cavaliere, 00133 Roma, Italy}

\altaffiltext{2}{Department of Astronomy and Astrophysics, 525
Davey Lab, Penn State University, University Park, PA 16802, USA}

\altaffiltext{3}{Harvard-Smithsonian Center for Astrophysics, 60
Garden St. Cambridge, MA 02138, USA}

\altaffiltext{4}{Osservatorio Astronomico Roma, Via Frascati 33,
00040 Monte Porzio Catone, Roma, Italy}

\altaffiltext{5}{NRAO, 1003 Lopezville Rd., Socorro, NM 87801,
USA}


\altaffiltext{6}{California Institute of Technology, Pasadena, CA,
91125, USA}

\altaffiltext{7}{UC Space Science Laboratory, Berkeley, CA
94720-7450, USA}

\altaffiltext{8}{Weizmann Institute, Rehovot 76100, Israel}

\begin{abstract}

We present  X-ray observations of the afterglow of GRB000926
performed around and after the break observed in the optical light
curve  two days after the burst. The steep X-ray light curve
observed around the break confirms the presence of this feature in
X-rays. However, the spectral and temporal properties are not
consistent with a standard jet scenario based on synchrotron
emission, requiring a more complicated model. We find that X-ray
and optical data are compatible with a moderately collimated
fireball (with opening angle $\theta\approx25 \deg$) expanding in
a dense medium ($n\approx 4\times 10^4$ cm$^{-3}$). This produces
two breaks in the light curve. The first, at $t\approx$ 2 days, is
due to  jet behavior. The second, around 5 days, is attributed to
the transition of the fireball to a non-relativistic expansion.
This transition predicts  a{\it flattening} of the light curve,
that explains the late X-ray measurement in excess above the
extrapolation of the simple jet scenario, and is also consistent
with optical data.

\end{abstract}


\section{Introduction}

Afterglows of Gamma Ray Bursts (GRB) are typically characterized
by a power law behavior  in  time and in frequency, i.e. the
observed flux $F\approx (t-t_0)^{-\delta} \nu^{-\alpha}$. This
property is nicely accounted for by the fireball model, in which
the afterglow is produced by the interaction of a relativistic
expanding shell  with an external medium (e.g. \cite{wrm}).
 The spectral
and temporal slopes depend on the geometry (e.g. spherical vs jet
(\cite{sari})) and on the density distribution of the external
medium (e.g. constant density -- hereafter ISM -- vs. wind
(\cite{chevalier})).
 A change of the slope (i.e. a break)  of the  light curve is produced
either when a spectral break (like that associated to $\nu_c$, the
cooling frequency of electrons ) transits  the observed frequency
window or when  ``bulk'' variations,- i.e. global changes of the
kinematics or hydrodynamics of the fireball,- take place. The
latter is the case of jet expansion or the transition from
relativistic to non-relativistic  expansion (hereafter named NRE).
In contrast to the effect of spectral breaks, jet expansion and
NRE produce achromatic breaks in the light curves. In  a jet this
happens when the beaming angle of the fireball ($\approx
\Gamma^{-1}$, where $\Gamma$ is the Lorentz factor), becomes
similar to the opening angle of the jet, $\theta$
(\cite{sari,roads}), at the time $t_{\theta}$:

\begin{equation}
t_{\theta}\approx (1+z)(\frac{E_{i,53}}{n})^{1/3}
(\frac{\theta}{0.1})^{8/3} {\rm days},
\end{equation}

where $n=n_6\ 10^{6} \rcm^{-3}$ is the density and $E_i= E_{i,53}\
10^{53}\ \rerg\ \rs^{-1}$  is the  energy the fireball would have
if it were spherically symmetric; i.e., in the case of a jet of
solid angle $\Delta\Omega$, its energy would be
$E=\frac{\Delta\Omega}{4\pi}E_i$.

 The transition to NRE instead
occurs  when the blast wave has swept up a rest-mass energy equal
to its initial energy, at a time (e.g. \cite{wrm}, Livio \& Waxman
(2000), hereafter LW):

\begin{equation}
t_{NRE} \approx 3 \frac{1+z}{2} (\frac{E_{53}}{n_6})^{1/3} {\rm
days}
\end{equation}

{\it Both} jet {\it and} NRE can therefore yield achromatic breaks
 a few days after the burst.
 The NRE requires an external medium of
a much higher density than that of a jet. This is indeed what one
would expect if GRB originate  in dense star forming regions, a
scenario for which supporting evidence is accumulating  (e.g.
\cite{gb991216,kulkarni}).

So far, evidence of achromatic breaks has been claimed in some
optical afterglows (e.g.\cite{gb990705_harrison,gb990123}), but an
{\it unambiguous} detection of this feature in the X-ray range is
still missing (e.g. \cite{gb990705,pian}). Most of the authors
attribute this feature to  jet behavior, although this model may
have some problem in accounting for the sharpness of the
transition (\cite{rf}). In fact, there is no - a priori - reason
to exclude a NRE and there is at least one case (GRB990123) in
which it has been attributed to either cases
(\cite{gb990123,dai}). In GRB000926 a break in the optical light
curve at $t\approx 2$ days has been
 reported by several groups (\cite{fynbo,veillet,halpern,rol},
 Price \etal (2000, hereafter P0).
The observed variation of the temporal slope of $\approx 2$ is
much larger than the value of 0.25 expected by a spectral break
(\cite{sari}). Indeed, a jet scenario has been proposed to explain
this behavior (\cite{price}).

 In this paper we report on BeppoSAX and  Chandra
observations of the afterglow of this burst (Sect. 2 \& 3). Those
observation occurred around and after the break observed in the
optical. We compare  X-ray with optical data
 (Sect.4) and
show that they are  explained by  jet behavior {\it and} a NRE
into a high density medium.

\section{Observations, data analysis and the X-ray spectrum}

The gamma-ray burst GRB000926 was detected by the Interplanetary
Network (IPN) on $t_0=\ $Sep.26.99 U.T. (\cite{ipn}). It was a
moderately bright GRB, with a revised 25-100 keV fluence of
$6.2\times 10^{-6} \fluence$, and a duration of about 25 s
(\cite{ipn}). Optical observations (\cite{dall}, \cite{gorosabel})
showed  an optical counterpart, at a very bright level ($R=19.3$)
one day after the GRB.  This is the second brightest afterglow
ever observed in the optical band. This finding triggered two
rapid target of opportunity  observations in X-rays, one  with
BeppoSAX and the other with Chandra. A second Chandra observation
was scheduled around $t=t_0+12$ days, to study the late time
evolution of the afterglow.

The BeppoSAX (\cite{psb,boella}) observation  started on Sep.
29.03 ($t-t_0=$2 days) and lasted for 12 hrs. Effective exposure
times were 20 ksec for the MECS(1.6-10 keV) and 5 ksec for the
LECS(0.1-10 keV). The analysis of the MECS image showed a
previously unknown fading source in a position consistent, within
the error of $50''$, with the optical transient, and therefore
identified as the X-ray afterglow of GRB000926
(\cite{piro1,piro2}). The source was only marginally detected in
the LECS. The average count rate of the source in the MECS(1.6-10
keV) is $(3.2\pm0.6)\times 10^{-3} cts/s$. The spectrum is well
fitted by  a power law with energy  index
$\alpha=(0.9\pm0.7)$\footnote{ Hereafter errors and upper limits
on fit parameters, including temporal slopes, correspond to 90\%
confidence level for one parameter of interest} and absorption by
our Galaxy ($N_H=2.7\times 10^{20} \rcm^{-2}$) and F(1.6-10
keV)=F(0.2-5 keV)$=(2.6\pm0.6)\times 10^{-13} \flux$. We have
checked different background subtraction methods, finding that the
corresponding systematic effect amount to less than about 15\%.
This has been included in the flux error quoted above.

A Chandra X-ray Observatory (\cite{chandra}) TOO observation
(hereafter C1) with the ACIS-S began on  Sep. 29.674 ($t-t_0$=2.7
days) and ended on Sep. 29.851, with an exposure time of 10ks. The
X-ray afterglow was observed at a position RA(2000)=17h 04m 09.6s;
Decl. (2000)= +51 47' 8.6" with an uncertainty of $2"$, consistent
with the optical position. The source count rate in the 0.3-5 keV
range is $(2.9\pm0.2)\times 10^{-2}$ cts s$^{-1}$, corresponding
to a flux F(0.2-5 keV)$=(1.2\pm1.0)\times 10^{-13} \flux$ for the
best fit power law spectrum. This has $\alpha=(0.8\pm0.2)$, i.e.
consistent with the BeppoSAX slope, and  absorption
$N_H=(5\pm3)\times 10^{20} \rcm^{-2}$, consistent with that of our
Galaxy. Since the BeppoSAX and Chandra spectral shapes are
consistent, we have fitted them simultaneously linking the
spectral parameters (but for the normalizations). We obtain a good
fit ($\csq=25.6, \nu=26$) with $\alpha=0.85\pm0.15$, a ratio of
the BeppoSAX over Chandra normalization of $2.3\pm0.6$ and some
marginal evidence of intrinsic absorption (in addition to the one
in our Galaxy) in the rest frame of the GRB (z=2.06
\cite{fynbo_z,castro})), corresponding to a column density
$N_{HGRB}=(0.4^{+0.35}_{-0.25})\times 10^{22} \cmm2$. The
residuals around 2 keV, i.e. the iron line region in the rest
frame of the GRB (Fig.1 ), are marginal, with an upper limit on an
iron line of $I<2\times 10^{-5} \cmm2$ s$^{-1}$, corresponding to
an equivalent width $EW<1$ keV. For comparison, the iron line
detected by Chandra in GRB991216 has $EW=0.5\ $ keV
(\cite{gb991216}).

\begin{figure}
\plotone{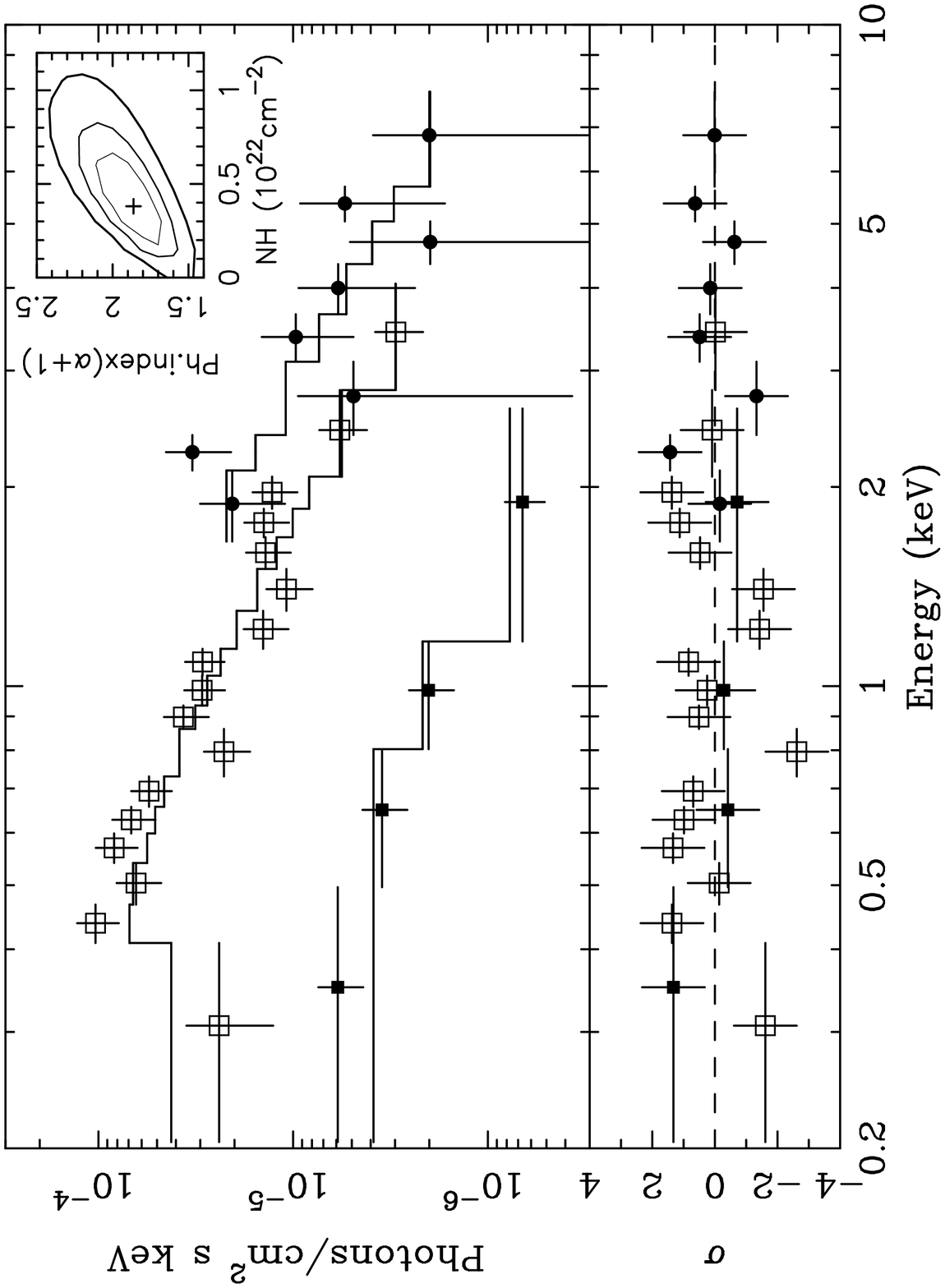} \caption{The X-ray spectrum of the afterglow of
GRB000926 as observed by the ACIS-S of Chandra (first observation:
open squares; second observation: filled squares ) and the MECS of
BeppoSAX (filled circles; the LECS is not shown for clarity). The
continuous lines represent the best fit power law convoluted
through instrument responses. Deviations from the best fit in
$\sigma$ are plotted in the lower panel. In the inset we show the
the contour plot of the photon index ($\alpha+1$) vs $N_H$ with
the 99\% (thick line) , 90\% (medium-thick line)  and 68\% (thin
line) confidence regions}

\label{spettro}
\end{figure}

The last Chandra observation (hereafter C2) took place on
Oct.10.176 and ended on Oct. 10.58. The afterglow count rate had
faded to $2.0 \pm 0.75 \times 10^{-3} {\rm c s^{-1}}$
(0.3--5.0~keV), allowing $\sim 60$~counts to be collected in the
33~ksec exposure.  With so few counts it is difficult to place
good constraints on ${\rm N_H}$ and $\alpha$ simultaneously, so we
fix $N_{HGRB} = 0\ ({\rm or}\ 0.4) \times 10^{22} cm^{-2}$ and
find $\alpha = 0.85 \pm 0.15\ {\rm(or}\ 1.2 \pm 0.3))$. The flux
in the 0.2-5 keV range derived by the best fit power law is equal
in both cases to $(8.3\pm1.1)\times 10^{-15} \flux$. An overall
fit on the three data sets (see Fig. \ref{spettro}) gives
$\csq/\nu=30.8/24$, with spectral parameters consistent with those
derived previously (see the inset in Fig.1) and  a ratio of the
normalizations C2/C1 of $(7.7\pm1.1)\times 10^{-2}$.

We find  marginal evidence of a steeper spectrum  in C2. The ratio
of count rates in the 0.2-1.5 keV range over 1.5-8 keV is in fact
$2.1\pm0.6$ and $3.9\pm0.2$ in C1 and C2 respectively. If we leave
the slope of C2 unlinked in the combined fit of all data sets, we
get $\alpha_{C2}=1.23\pm0.3$. While this result is of marginal
statistical significance, it is interesting to note that the
difference in slope between C2 and C1+BeppoSAX  is
$\Delta\alpha=0.4\pm0.3$, i.e. consistent with the steepening of
1/2 expected when $\nu_c$ passes into the energy window.


An intrinsic column density $N_{HGRB}\approx 0.4\times 10^{22}
\cmm2$ is marginally detected at a confidence level of $\approx
98\%$ (see the inset in Fig.1). We remark that, since most of the
local gas is ionized by GRB photons (\cite{boettcher}), the real
column density can be much higher.
Assuming that the relation between the optical extinction
$A_V$ and $N_H$ is similar to that in our own galaxy (e.g.
\cite{predehl}),  gives $A_v\approx 2$. This value is $\approx
2-20$ greater than that derived by optical measurements
(\cite{price}), implying a dust-to-gas ratio much lower than that
of our Galaxy. This effect has been observed in other GRB
(\cite{galama_dust}) and it may be the result of destruction of
dust by the GRB radiation (\cite{waxman_dust}).

\section{A very  steep X-ray decay}

\begin{figure}
\plotone{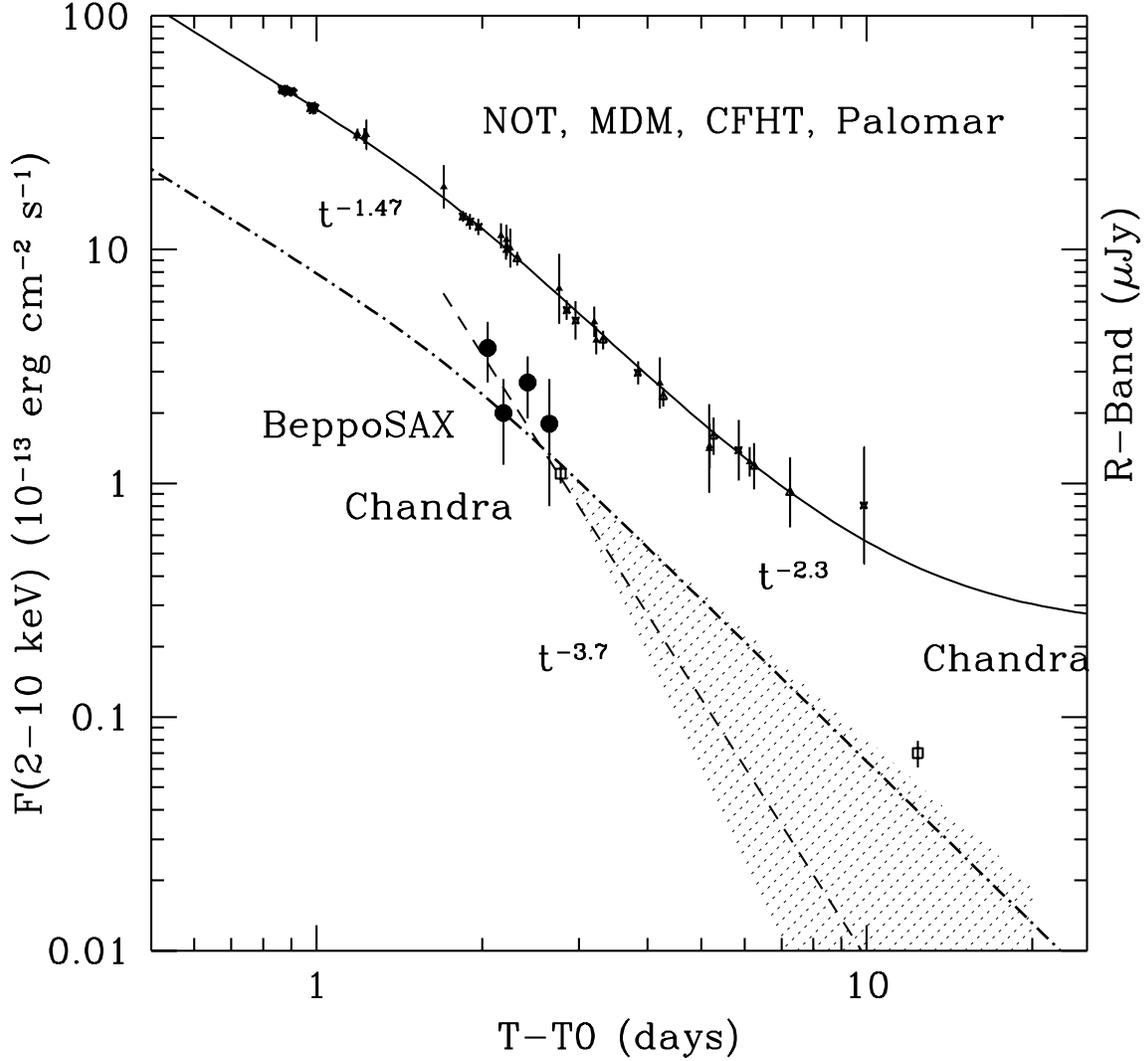} \caption{ The X-ray and optical (R band) light
curves of the afterglow of GRB000926. Optical data are from Dall
\etal (2000), Fynbo \etal (2000a, 2000b), Halpern \etal (2000),
Price \etal (2000), Veillet \etal (2000). The X-ray light curves
have been derived by converting the flux at 1 keV, which is not
significantly dependent on the spectral shape, to the F(2-10) keV
flux, by adopting the best fit power law with $\alpha=0.85$ for
all the spectra. The dashed line is the best fit to the data of
the BeppoSAX and first Chandra observations. The allowed region
(at $90\%$ confidence level) obtained by extrapolating the
previous fit is identified with the shaded area. The continuous
line is the best fit to the optical data including the effect of
the host galaxy (P0). The dash-dot line is the template of the
optical afterglow (obtained from the best fit to the data
subtracting the galaxy) normalized to the X-ray data at $T-T_0<3\
$days. } \label{curve}
\end{figure}

 In the BeppoSAX data the source exhibits a
substantial decay, decreasing by a factor of $(1.7\pm0.5)$ in 6
hours. This corresponds to a power law decay $F\propto
(t-t_0)^{-\delta_X}$ with slope $\delta_X\approx4$. This behaviour
is consistent with  the first Chandra observation, that took place
near the end of the BeppoSAX observation, and whose flux was a
factor $2.3\pm0.6$ times lower than the average BeppoSAX flux.
Combining these two data sets (Fig.2) we derive a slope $\delta_X=
3.7^{+1.3}_{-1.6}$, in agreement with the optical slope measured
after the break.
The late Chandra point appears to deviate from this law, the slope
connecting C1 with C2 being $\delta_x=1.70\pm0.16$. However, when
the large error of the earlier slope is considered, the two
measurements are nearly compatible (Fig. \ref{curve}). A combined
fit to all the X-ray data points gives then
$\delta_X=1.89^{+0.19}_{-0.16}$ with F(2-10
keV)=$1.19\times10^{-12} \flux$ at $t=1\ $ day.

The X-ray decay of GRB000926  is much steeper than that typically
observed in X-ray afterglows, where $\delta_X\approx1.4$
(\cite{piro_bo}). We remark, however, that previous observations
covered the evolution of the afterglow primarily at $t-t_0<2$
days. The data presented here are  instead sampling the X-ray
behaviour at later times. It is in this domain that we expect
deviations due to either a jet or NRE.

\section{Jet vs NRE of a spherical fireball in a dense medium}

The optical data (Fig.2) are nicely fitted by the standard jet
model with synchrotron emission (\cite{price}) but, when X-ray
data are considered, we find inconsistencies with this scenario.
The relationship linking the temporal and spectral slopes are
(\cite{sari}) $\delta=2\alpha+1$ ($\nu<\nu_c$) and
$\delta=2\alpha=$ ($\nu>\nu_c$). We find then that the X-ray
temporal and spectral slopes $\delta_X=1.9$ and $\alpha_x=0.85$
are consistent  only with the case of a cooling frequency below
the X-ray range. The slope $p$ of the electron distribution
$N_e\propto \gamma^{-p}$ would then be
$p=\delta=2\alpha=1.7\pm0.3$, i.e. marginally compatible with the
minimum slope needed to avoid an infinite energy content in
electrons.
 A more relevant problem of this model  is that the predicted
 temporal slope is significantly lower
 than the one measured in optical (\cite{price}; see
 also Fig. \ref{spettro}), $\delta_O=2.3\pm0.08$.
Furthermore, the electron slope derived from the optical data
under the same scenario  ($p=2.39\pm0.15$) is not consistent with
the X-ray derivation. Finally we note that the - admittedly
marginal - evidence of a spectral steepening in C2, argues against
a jet behavior, during which $\nu_c$ remains constant.

We now examine the  alternative explanation of a transition to NRE
of a spherical fireball . The relationships linking $\delta$ and
$\alpha$
 are given in Livio \& Waxman (2000). We find that the X-ray data are
consistent with a NRE either in a uniform density medium ( which
gives $\delta=2.17\pm0.45$) or in a wind ($\delta=2.3\pm0.35$).
The cooling frequency is now {\it above} the X-ray range, and
therefore $p=2.7\pm0.3$.  In this case  the optical and X-ray
spectral data points should lie on a single power law. We have
evaluated the slope
$\alpha_{OX}=-\frac{log(F_X/F_O)}{log(\nu_X/\nu_O)}$ of the power
law connecting the flux at 1 keV observed by Chandra ($F_X$) with
the R band flux extrapolated at the time of the Chandra
observation ($F_O$, Fig.2) and corrected for the extinction
(\cite{price}). We find $\alpha_{OX}=0.9-1.1$, in agreement with
the slope measured in X-rays.
Before the break the evolution is indistinguishable from that of a
spherical relativistic fireball, which follows a decay with
$\delta=1.3\pm0.2$ (ISM) or $\delta=1.8\pm0.2$ (wind). The latter
case is not compatible with the optical slope measured before the
break (\cite{price}), and we conclude that  both optical and X-ray
data are consistent with a spherical expansion in a uniform
medium, with the temporal break produced by a transition to NRE.
From eq.(2) we derive $n_6\approx 10$. We have estimated,
following Freedman and Waxman (2000), a fireball energy comparable
to the observed $\gamma$-ray isotropic energy, which is
$E_{i,\gamma}=8\times 10^{52}$erg ($z=2.03$ (\cite{castro}), $H_0=
 75\ $ km/s/Mpc, $q_0=1/2$).
The absorbing column density is $N_H\approx nr_{NRE}\approx\
10^{23}\ \rcm^{-2}$, where the radius $r_{NRE}\approx 10^{16}$cm
(\cite{livio}). This is in apparent disagreement with the upper
limits derived by the X-ray spectrum. This disagreement is
rectified by recalling that the medium is heavily ionized by the
GRB photons, and therefore its effective absorption is negligible
(\cite{boettcher}).


\section{A collimated fireball in a dense medium }

There remains, however, a discrepancy between optical and X-ray
data, that is not accounted for by  previous models. The  X-ray
decay  {\it after} the break ($\delta_X=1.70\pm0.16$) is
significantly flatter than the optical one
($\delta_O=2.30\pm0.08$, \cite{price}). In fact, the extrapolation
of the X-ray flux with the optical slope at the time of the second
Chandra observation underestimates the observed flux by a factor
of $2\pm0.5$ (Fig.\ref{curve}).  This indicates the presence of a
{\it flattening} taking place in between the two sets of
observations, as   predicted in the case of a jet expansion
followed by a transition to NRE(\cite{livio}). In this case,
 the lateral expansion of the jet
begins when it is relativistic, and during the lateral expansion
stage there is rapid decline of the flux due to the rapid decrease
of the fireball energy per unit solid angle (as this angle
increases). This is  producing the first break in the light curve.
The time at which the jet completes its side-ways expansion, is
roughly the same time at which the flow becomes sub-relativistic.
At this stage, lateral expansion is no longer important, and the
rapid flux decline due to lateral expansion ceases, leading to a
flattening of the light curve.
 This effect appears more
evident in X-rays, which are not contaminated at later times by
the presence of a fairly bright host galaxy (\cite{price}).

We show now that both optical and X-ray data are in agreement with
this scenario. The light curve of the afterglow is composed by
three subsequent power laws with decay slopes $\delta_1$,
$\delta_2$, and $\delta_3$, that  we have modeled with the
empirical formula :

\begin{equation}
F(t)= F(t_{b1})\ \frac{ [ (\frac{t}{t_{b1}})^{\delta_1 s}
+(\frac{t}{t_{b1}})^{\delta_2 s}]^{-1/s}
 }{[1+(\frac{t}{t_{b2}})^{\delta_2 s}]^{-1/s}[1+(\frac{t}{t_{b2}})^{\delta_3 s}]^{1/s}}
\end{equation}

The parameter $s$ describes the sharpness of the breaks. The
temporal slopes are related to the power law index of electrons by
relationships that depend on the location of $\nu_c$ and on the
density distribution of the external medium (\cite{livio}). We can
 exclude most of the cases as follows.
$\alpha_{OX}\approx\alpha_x$ implies that $\nu_c$ is either below
the optical or above the X-ray range. In the first case, we would
have $p=2\alpha_X=1.7$ that, besides being less than 2, would give
temporal slopes  flatter than observed. In fact:
$\delta_1=\frac{(3p-2)}{4}=0.8$, $\delta_2=p=1.7$  and
$\delta_3=1+\frac{7}{6}(p-2)=0.65$ (wind) or
$\delta_3=1+\frac{3}{2}(p-2)=0.55$ (ISM).  During the jet phase
$\nu_c=const$ while, in the NRE phase,  $\nu_c$ increases
(decreases) with time for a wind (ISM).  In the case of a wind
$\nu_c$ could move above the optical range, but  still giving a
flat slope $\delta_3=\frac{3}{2}+\frac{7}{6}(p-2)=1.15$.

\begin{figure} \plotone{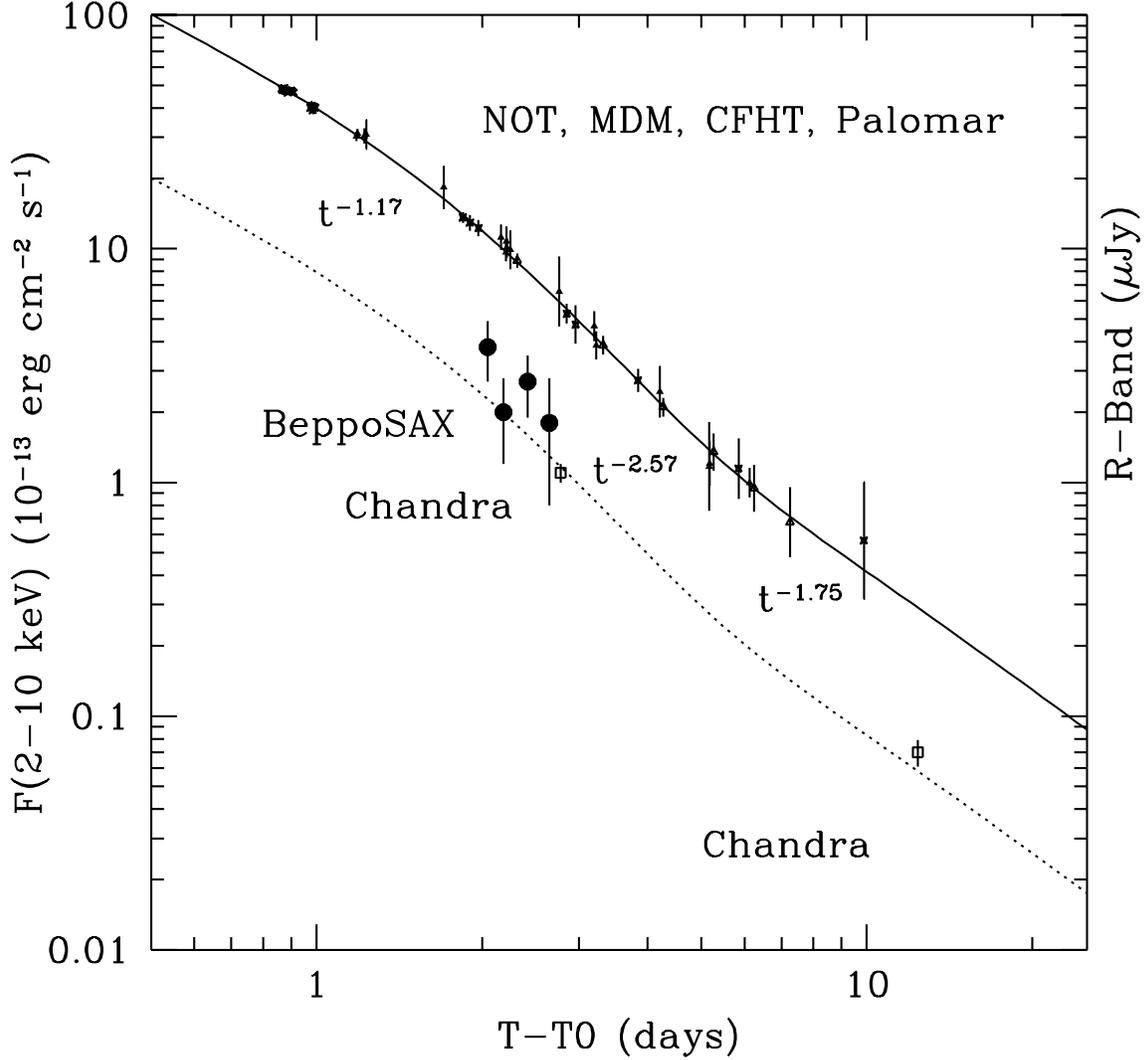} \caption{ Same as
in Fig.2 but with the contribution of the  galaxy subtracted from
optical data. The continuous and dotted light curves represent the
expected behaviour in the case of a moderately collimated jet
 with a transition to non-relativistic expansion at t=5.5
days. Note the flattening of light curves both in the optical and
X-rays after this second break.} \label{curve2}
\end{figure}

Let us now examine the case of $\nu_c$ lying above the X-ray
range. This implies $p=2\alpha_X+1=2.7$. The case of a wind is
excluded because $\delta_1=1.8$, i.e. greater than observed. For a
ISM we have $\delta_1=\frac{3}{4}(p-1)=1.3$, $\delta_2=p=2.7$ and
 $\delta_3=\frac{9}{10}+\frac{3}{2}(p-2)=1.95$, which
 appear to be consistent with the data. We have  assessed this
 case more quantitatively by fitting  eq.(3),
 with the slopes expressed as a function of $p$, to the
 optical  data (with the contribution of the host galaxy
subtracted) . We obtain $p=2.57\pm0.3$, in nice agreement with the
value derived from the X-ray spectra, $t_{b1}=1.76^{+0.7}_{-0.4}\
$ days, $t_{b2}=5.5\pm2\ $ days, and $s=1.7\pm0.3$. Moreover, this
same law (i.e. rescaling the normalization but with the same
parameters determined by the optical light curve) fits fairly well
all the X-ray data points, and in particular reproduce C2 (Fig.
\ref{curve2}). We also note that this law predicts a temporal
slope of about 2.6 around 2 days, nearer to the steep decay
observed in the BeppoSAX an C1 observations.


 The simultaneous determination of
$t_{\theta}=t_{b1}$ and $t_{NRE}=t_{b2}$ allows an unambiguous
determination (i.e. not depending on estimates of $E_i$ and $n$)
of the jet angle (\cite{livio}). In fact $t_{NRE}=(1+z)^{3/4}
(\frac{E_{i,53}}{n_6})^{1/4} t_{\theta}^{1/4}$ which, when
combined with eq.(1), gives
 $\theta=0.7(\frac{t_\theta}{t_{NRE}})^{1/2}=0.4$.
The  total $\gamma$-ray energy is then, for a two-sided jet,
$E_{\gamma}=0.16E_{\gamma,i}=1.2\times 10^{52} \rerg\ \rs^{-1}$.
Finally, the density $n_6=(1+z)^3 E_{i,53}
\frac{t_\theta}{t_{NRE}^4}\approx 0.04$. Note that the jet angle
is rather insensitive to the errors in the determination of break
times. Conversely, the density should be considered as an order of
magnitude estimate, due to the strong dependence on $t_{NRE}$ and
to model details (\cite{livio}). We have finally checked whether
the value of density is compatible with the broad band spectrum of
the GRB, following the prescriptions of Wijers \& Galama (1999).
Since the parameters one derives are dependent on the model
adopted, we have restricted the analysis to the data before the
break, when the fireball evolution is indistinguishable from the
spherical case.  From the radio data (Frail, private
communication; Harrison \etal (2001)) and the extrapolation of the
optical-to-X-ray spectrum at lower frequencies, we derive that the
synchrotron self-absorption frequency $\nu_a$ and frequency of the
maximum flux $\nu_m$ should be in the range between
$10^{10}-10^{12}$Hz, and that the flux at $\nu_m$ should be
$\approx\ $a few mJy. Furthermore, as we have shown before,
$\nu_c\gtrsim10^{18}$ Hz. The large range spanned by these
quantities gives  a density  spread over orders of magnitude that,
however, comfortably include the density estimate derived before.

\section{Conclusions}

In this paper we have shown that X-ray observations of the
afterglow of GRB000926 around and after the break observed in the
optical confirm the presence of this feature in X-rays as well.
However, they are not consistent with the  standard jet scenario
based on synchrotron emission, requiring a more complicated model.
We find that both the spectral and temporal properties of X-ray
and optical data are  compatible with a moderately collimated
fireball ($\theta\approx25 \deg$) expanding in a dense medium
($n\approx 4\times 10^4$ cm$^{-3}$). This produces two breaks in
the light curve. The first is due, as usual, to a jet behavior.
The second, around 5 days, is attributed to a transition to NRE
due to the high density of the external medium, and is followed by
a {\it flattening} of the curve, that accounts for the late X-ray
measurement, and is also preferred by optical data. Optical
measurements at later times should help to characterize this
behavior but, to do so, it will be necessary to actually single
out the afterglow from the host in the image. Current estimates of
the contribution of the host galaxy from the flattening exhibited
by the light curve will, in fact, cancel out any intrinsic
flattening due to the afterglow. Finally, we note that we cannot
exclude other possibilities like, for example, an event of
gravitational lensing (\cite{mao}) that amplified the flux around
12 days, or a contribution of an Inverse Compton component to the
late X-ray emission (Harrison \etal 2001). An analysis of
different options goes, however, beyond the scope of this paper.

Our result supports the association of GRB with star forming
regions. In fact, the density is consistent with that of very
dense molecular clouds ($n\approx 10^4$ cm$^{-3}$).  On the other
hand it is not as high as that required to produce the iron lines
observed in some X-ray afterglows (e.g.\cite{antonelli};
\cite{gb991216} and references therein), which is possibly
connected with much denser ($n\approx 10^{10}$cm$^{-3}$)
progenitor ejecta. In these cases the environment must be composed
by at least two regions, since the fireball requires a medium with
a density $n\ltsim10^7$cm$^{-3}$ to produce an afterglow lasting
at least a few days. The properties of this medium, in particular
the density, are not the same in all GRB. While in this GRB we
find $n\approx 4\times 10^4$cm$^{-3}$, in several other GRB the
density should be much lower, as indicated, for example, by the
absence of breaks in the optical light curves (\cite{kulkarni}) or
by the evidence of relativistic expansion  several weeks after the
GRB (\cite{wkf}). Another element of diversity in GRB appears to
be the magnetic field strength.  In this burst the  magnetic field
density is estimated to be less than $\approx 10^{-6}$ times the
equipartition value, comparable to that of GRB971214 (\cite{wg})
and GRB990123 (\cite{galama990123}), but four orders of magnitude
weaker than in GRB970508 (\cite{wg}). As suggested by Galama \etal
(1999), such differences in field strength may reflect differences
in energy flow from the central engine.

\begin{acknowledgments}

We thank the BeppoSAX \& Chandra teams for the support, in
particular H. Tananbaum for the organization of the Chandra TOO.
BeppoSAX is a program of the Italian space agency (ASI) with
participation of the Dutch space agency (NIVR). KH is grateful for
Ulysses support under JPL contract NAG 5-9503, and for NEAR
support under NASA grant NAG5-95903. GG \& PM acknowledge support
under NASA grant G00-1010X. MRG acknowledges support under NASA
contract NAG8-39073 to the Chandra X-Ray Center.

\end{acknowledgments}

\end{document}